\documentclass[12pt]{article}

\usepackage{amsmath}
\usepackage{amssymb}
\usepackage{amsthm}
\usepackage{booktabs}
\usepackage{graphicx}
\usepackage{natbib}
\usepackage{geometry}
\usepackage{setspace}
\usepackage{pdfpages}
\def\top{{\mathrm{\scriptscriptstyle T} }}
\DeclareMathOperator{\var}{var}
\DeclareMathOperator{\cov}{cov}
\newtheorem{theorem}{Theorem}
\newtheorem{assumption}{Assumption}
\newtheorem{remark}{Remark}
\allowdisplaybreaks[4]

\title{\bf Doubly Robust Estimation of Treatment Effects in Staggered Difference-in-Differences with Time-Varying Covariates}

\author{Yuhao Deng$^1$ and Le Kang$^{2,*}$ \\
{\small 1 Fred Hutchinson Cancer Center; 2 Shanghai University of Finance and Economics} \\
{\small Correspondence: Le Kang, lekang@sufe.edu.cn} \\
{\small 777 Guoding Road, Shanghai 200433, China}}

\date{}

\begin{document}

\maketitle

\begin{abstract}
With staggered adoption, difference-in-differences (DiD) analyses define the average treatment effect on the treated (ATT) at the group-period level and aggregate it into groupwise, periodwise, dynamic, and overall effects. We study a period-by-period version of conditional parallel trends, with covariates unaffected by treatment while allowing covariates to shift the trends in untreated potential outcomes. We propose an augmented inverse probability weighting (AIPW) estimator and an augmented inverse variance weighting (AIVW) generalization. Both are doubly robust and asymptotically normal, with closed-form influence-function-based inference for all group-period and aggregated ATTs. AIVW is statistically efficient in some special cases, while AIPW is computationally more efficient. We illustrate the methods with the staggered provincial adoption of a parallel college admission mechanism in China and its effect on justified envy. \par
\end{abstract}

\section{Introduction}

Difference-in-differences (DiD) is a widely used quasi-experimental method for estimating treatment effects from observational panel data by comparing changes in outcomes over time between treated and control units. With staggered adoption, treatment effects are commonly defined and estimated at the group-period level and then aggregated into groupwise, periodwise, dynamic, and overall average treatment effects on the treated (ATT) \citep{callaway2021difference, athey2022design}. This group-period formulation avoids the negative-weighting problems of two-way fixed effects estimators under heterogeneous treatment effects and keeps the target estimands model-free \citep{goodman2021difference, de2020two}. Because all aggregated ATTs are identifiable linear combinations of group-period ATTs, accurately and efficiently estimating the group-period ATTs is the central statistical problem in this literature.

Under staggered adoption, many not-yet-treated groups are typically available as controls at any given period. We build on the framework of \citet{callaway2021difference}. Their doubly robust estimators compare the long difference $Y_t - Y_{g-1}$ in group $g$ with the same long difference in a control group consisting of either the never-treated units or the units not yet treated by period $t$, conditioning on pre-treatment covariates. \citet{chen2025efficient} further derive the efficient estimator under parallel trends conditional on such baseline covariates by combining comparisons across all pre-treatment baseline periods and control groups. However, these works do not consider time-varying covariates.

In many applications, the covariates that plausibly shift untreated trends vary over time. \citet{caetano2022difference} and \citet{caetano2024difference} study DiD with parallel trends conditional on time-varying and time-invariant covariates. They show that two-way fixed effects regressions that include such covariates can be biased even when parallel trends hold conditional on them, discuss the bad-control problem that arises when covariates are affected by treatment, and propose regression adjustment, inverse probability weighting, and doubly robust estimators that condition the long difference $Y_t - Y_{g-1}$ on the time-invariant covariates and on the time-varying covariates in periods $g-1$ and $t$. Our paper is closest to this work; we restrict attention to time-varying covariates unaffected by treatment.

We study a different formulation of parallel trends. Our Assumption restricts adjacent-period changes: given the covariates $X_k$ of period $k$, which can include time-invariant covariates and current and lagged values of time-varying covariates, the mean untreated change $\Delta Y_k(\infty)$ does not depend on the adoption group. The long-difference assumption in \citet{caetano2024difference} instead restricts $Y_t(\infty)-Y_{g-1}(\infty)$ given the covariates at its two endpoints. Neither implies the other. If the untreated change in each period depends on the covariates of that period, the long difference depends on the intermediate covariate path, and conditioning on the endpoints does not remove this dependence when groups with the same endpoints follow different paths in between; conversely, the long-difference assumption places no restriction on intermediate periods. Our identification formula builds the counterfactual trend from adjacent-period changes, shifting the covariate distribution from the control groups to the treated group in each period, and uses every group not yet treated in each period. We combine these groups with a normalized inverse-conditional-variance rule, analogous to inverse variance weighting (IVW) in meta-analysis \citep{lee2016comparison}, which is the generalized least squares combination within the period.

Our contributions are as follows. (i) Under period-by-period conditional parallel trends and covariate exogeneity, we give an identification result and doubly robust AIPW and AIVW estimators of group-period ATTs and their aggregations. They are consistent when parallel trends fail conditional on baseline covariates alone. (ii) The AIVW weights are the generalized least squares combination of the group-specific trend moments within each period, and we give conditions under which AIVW attains the semiparametric efficiency bound; outside these conditions, we make no efficiency claim. (iii) We provide closed-form influence-function-based inference for all ATTs; the overall ATT of AIPW is a weighted sum of outcome-regression residuals. (iv) Our simulations separate the sources of estimation precision. Conditioning on a time-varying covariate that predicts the untreated trend makes AIPW substantially more precise than estimators that adjust only for baseline covariates; AIVW improves on AIPW only when conditional variances differ across not-yet-treated groups and outcome changes are close to serially independent. The proposed estimators are preferable when the untreated change in each period depends on the covariates of that period.

The remainder of this paper is organized as follows. Section \ref{sec2} defines the estimands and lists the identification assumptions. Section \ref{sec3} presents the estimators and their properties, including doubly robustness and asymptotic normality. Section \ref{sec4} reports summary simulation findings, while details are deferred to the supplementary material. Section \ref{sec5} studies the effect of a parallel college admission mechanism on justified envy using staggered provincial reforms in the China National College Entrance Examination. Section \ref{sec6} concludes.

\section{Causal difference-in-differences} \label{sec2}

\subsection{Average treatment effects on the treated}

In the staggered difference-in-differences setting, suppose there are $T+1$ periods. In the pre-treatment period ($t=0$), none of the units is treated. A unit may initiate treatment at any period $t \in \{1, \ldots, T\}$ after the pre-treatment period or never receive treatment throughout the experiment. Let $D_t$ be the treatment received in period $t$. We denote $G=g$ if the unit initiates treatment in period $g$, i.e., $D_0 = \cdots = D_{g-1} = 0$ and $D_g = 1$. Treatment is absorbing, so $D_t = I(G \leq t)$; carryover and delayed effects are allowed. Never-treated units have $G = \infty$, coded as $G=T+1$ in computations. For each unit, let $Y_t(g)$ denote the potential outcome in period $t$ if treatment were initiated in period $g$. We denote the potential outcome of never being treated as $Y_t(\infty)$. The unit treatment effect in period $t$ if it were assigned to group $g \leq t$ is $Y_t(g) - Y_t(\infty)$.

Because $Y_t(g)$ and $Y_t(\infty)$ are never observed for the same unit, we target the average treatment effect on the treated (ATT) in group $g$ and period $t$, for $1 \leq g \leq t \leq T$,
\[
\tau_{g,t} = E\{Y_t(g)-Y_t(\infty) \mid G=g\}.
\]

\begin{figure}
    \centering
    \includegraphics[width=0.8\textwidth]{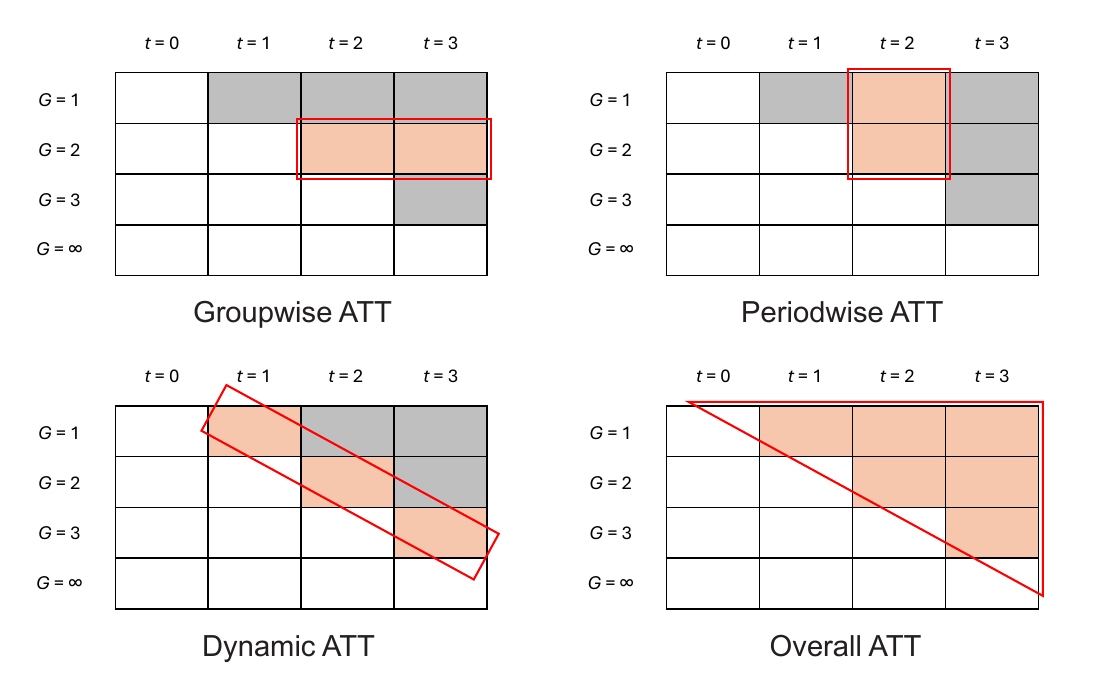}
    \caption{Groupwise ATT, periodwise ATT, dynamic ATT, and overall ATT aggregated from group-period ATTs.} \label{fig1}
\end{figure}

Note that group-period ATTs are only defined in a triangular array. To aggregate group-period ATTs, we define the ATT in group $g$ as
\[
\tau_g^{\text{grp}} = \frac{1}{T-g+1} \sum_{t=g}^{T} \tau_{g,t},
\]
the ATT in period $t$ as
\[
\tau_t^{\text{prd}} = \frac{1}{\sum_{g=1}^{t} P(G=g)} \sum_{g=1}^{t} P(G=g) \tau_{g,t},
\]
the dynamic ATT of duration $s$ as
\[
\tau_s^{\text{dyn}} = \frac{1}{\sum_{g=1}^{T-s+1} P(G=g)} \sum_{g=1}^{T-s+1} P(G=g) \tau_{g,g+s-1},
\]
and the overall ATT as
\[
\tau = \frac{1}{\sum_{g=1}^{T} (T-g+1)P(G=g)} \sum_{g=1}^{T}\sum_{t=g}^{T} P(G=g) \tau_{g,t}.
\]
In general, these aggregated ATTs are in the form of
\[
\frac{1}{\sum_{(g,t): (g,t)\in\mathcal{S}, g \leq t}P(G=g)} \sum_{(g,t): (g,t)\in\mathcal{S}, g \leq t}P(G=g) \tau_{g,t},
\]
where $\mathcal{S}$ is the target population composed of eligible treated cells $(g,t)$ that satisfy a specific criterion. In the example of Figure \ref{fig1}, with periods $t=0,\ldots,3$ and groups $G\in\{1,2,3,\infty\}$, the target population in the groupwise ATT $\tau_2^{\text{grp}}$ is $\mathcal{S}_{g=2} = \{G=2\}$, the target population in the periodwise ATT $\tau_2^{\text{prd}}$ is $\mathcal{S}_{t=2} = \{G\leq2\}$, and the target population in the dynamic ATT $\tau_1^{\text{dyn}}$ is $\mathcal{S}_{g=t} = \{G\leq3\}$. The overall ATT is defined on all treated unit-periods, $\mathcal{S}_{g\leq t} = \{D_t = 1: t=1, \ldots, T\}$.

Since time-varying covariates are measured after treatment, we write the time-varying covariates $X_t(g)$ in period $t$ as a potential outcome of group assignment $g$. If no treatment is received, the never-treated time-varying covariate path is $X_t(\infty)$. Let $X_t$ be the observed time-varying covariates in period $t$, which may include baseline covariates. The observed data in the sample $\{O_i = (G_i,X_{0i},Y_{0i},\ldots,X_{Ti},Y_{Ti}): i=1,\ldots,n\}$ consist of $n$ independent and identically distributed units with observation $O = (G, X_0, Y_0, \ldots, X_T, Y_T)$.

\subsection{Identification}

We define the change in potential outcomes in period $t$ as $\Delta Y_t(g) = Y_t(g) - Y_{t-1}(g)$. We assume no anticipation as follows.

\begin{assumption}[No anticipation] \label{ass1}
    $Y_t(g) = Y_t(\infty)$ almost surely for every $t<g$.
\end{assumption}

Under no anticipation, the untreated potential outcome before adoption does not depend on when treatment starts. The distribution of $Y_t(\infty)$ may still differ across groups, even conditional on observed covariates, because of unmeasured confounding.

\begin{assumption}[Covariate exogeneity to treatment] \label{ass2}
    $X_t(g)=X_t(\infty)$ almost surely for every $g\in\{1,\ldots,T,\infty\}$ and $t\in\{0,\ldots,T\}$.
\end{assumption}

We assume parallel trends conditional on the untreated covariates.

\begin{assumption}[Parallel trends] \label{ass3}
   $E\{\Delta Y_t(\infty) \mid X_t(\infty), G=g\} = E\{\Delta Y_t(\infty) \mid X_t(\infty), G=\infty\}$ for every $g\in\{1,\ldots,T,\infty\}$ and $t\in\{1,\ldots,T\}$.
\end{assumption}

Assumption \ref{ass3} is not implied by unconfoundedness of $G$ given baseline covariates, because conditioning on post-baseline covariates can reintroduce dependence on $G$. If $X_t(\infty) \equiv X$ is time-invariant, it implies that $E\{Y_t(\infty) - Y_s(\infty) \mid X, G\}$ does not depend on $G$ for any $s,t$. The covariates $X_t(\infty)$ can contain current values, lags, or other summaries of the history up to period $t$. For example, let $Z$ be a baseline covariate and $Z_t$ a time-varying covariate \citep{caetano2024difference}, and consider the structural model
\[
Y_t(\infty) = f_t(Z, Z_t) + U + \varepsilon_t,
\]
where $U$ is a time-invariant unmeasured confounder that may influence treatment assignment and $E(\varepsilon_t-\varepsilon_{t-1}\mid Z,Z_t,Z_{t-1},G)=0$. The change $\Delta Y_t(\infty) = f_t(Z,Z_t) - f_{t-1}(Z,Z_{t-1}) + \varepsilon_t - \varepsilon_{t-1}$ does not involve $U$, so Assumption \ref{ass3} holds with $X_t(\infty) = (Z, Z_t, Z_{t-1})$. In this model, the long-difference assumption conditional on $(Z,Z_{g-1},Z_t)$ also holds, because $Y_t(\infty)-Y_{g-1}(\infty)$ depends on the covariates only through their values in periods $g-1$ and $t$. The two assumptions differ when the untreated change depends on the level of the covariate, for example $\Delta Y_t(\infty)=h_t(Z,Z_t)+\varepsilon_t-\varepsilon_{t-1}$ with $E(\varepsilon_t-\varepsilon_{t-1}\mid Z,Z_t,G)=0$. Assumption \ref{ass3} then holds with $X_t=(Z,Z_t)$, whereas
\[
Y_t(\infty)-Y_{g-1}(\infty)=\sum_{k=g}^{t}h_k(Z,Z_k)+\varepsilon_t-\varepsilon_{g-1}
\]
depends on the intermediate values $Z_g,\ldots,Z_{t-1}$, and conditioning on the endpoint covariates removes this dependence only if the conditional mean of $\sum_{k=g}^{t-1}h_k(Z,Z_k)$ given $(Z,Z_{g-1},Z_t)$ is the same across groups. Conversely, the long-difference assumption restricts only the long differences, so it can hold when Assumption \ref{ass3} fails in intermediate periods.

\begin{assumption}[Positivity] \label{ass4}
    There exists a positive constant $\eta>0$ such that $P(G=g \mid X_t(\infty)) > \eta$ for every $g\in\{1,\ldots,T,\infty\}$ and $t\in\{0,\ldots,T\}$ with probability 1.
\end{assumption}
\begin{assumption}[Consistency] \label{ass5}
    $X_t = X_t(G)$ and $Y_t = Y_t(G) = \sum_{g\in\{1,\ldots,T,\infty\}}I(G=g)Y_t(g)$.
\end{assumption}

Positivity ensures that, given $X_t$, every group is represented, so that trends can be borrowed from untreated groups without extrapolation. With time-varying covariates, the covariate distribution must instead be shifted from the control groups to group $g$ in each period.

\begin{theorem}
Under Assumptions \ref{ass1}--\ref{ass5}, the group-period ATT is identifiable for every $1\leq g \leq t \leq T$,
\begin{equation}
\tau_{g,t} = E(Y_t - Y_{g-1} \mid G=g) - \sum_{k=g}^{t} E\{E(Y_k-Y_{k-1} \mid X_k, G=\infty) \mid G=g\}. 
\end{equation}
\end{theorem}

The proof is given in Supplementary Material A.1. This identification result is related to the build-the-trend estimators discussed in \citet{marcus2021role}. By contrast, under the long-difference assumption of \citet{caetano2022difference} and \citet{caetano2024difference}, with $Z$ and $Z_t$ as in the example above, $\tau_{g,t}=E(Y_t-Y_{g-1}\mid G=g)-E\{E(Y_t-Y_{g-1}\mid Z,Z_{g-1},Z_t,G=\infty)\mid G=g\}$, a single regression of the long difference on the endpoint covariates. The two formulas coincide for $t=g$ when $X_g$ contains the covariates of periods $g-1$ and $g$; for $t>g$, Theorem 1 instead averages each adjacent-period regression over the period-$k$ covariate distribution of group $g$ and can use every group not yet treated at $k$.

\subsection{Choice of time-varying covariates}

The choice of $X_t$ affects both bias and variance of the estimator \citep{caetano2022difference, caetano2024difference}. First, $X_t$ should include covariates that shift the untreated trend, such as $Z_t$ in the structural model above; omitting one whose distribution differs across groups violates Assumption \ref{ass3}. Second, these covariates must not respond to treatment, $X_t(g)=X_t(\infty)$; otherwise, conditioning on the observed $X_t(g)$ in treated groups does not recover the counterfactual trend, the bad-control problem discussed by \citet{caetano2022difference}. Section \ref{sec5} argues that the time-varying covariates of our application satisfy this restriction. Third, because Theorem 1 conditions on $X_k$ in every period, covariates that are only weakly related to the trend enlarge the conditioning set and make positivity (Assumption \ref{ass4}) harder to satisfy without reducing bias, so a parsimonious $X_t$ is preferable.

\section{Estimation} \label{sec3}

\subsection{Proposed estimators}

Let $\mathbb{P}(h) = \int h(o)\,dP(o)$ denote expectation under the true distribution $P$ of the observed data, and $\mathbb{P}_n(h) = n^{-1}\sum_{i=1}^{n}h(O_i)$ the empirical measure. Let $\pi_{g,t}(X_t) = P(G=g \mid X_t)$ be the generalized propensity score, $\mu_{g,t}(X_t) = E(Y_t \mid X_t, G=g)$ the mean outcome, and $\delta_{g,t}(X_t) = E(\Delta Y_t \mid X_t, G=g)$ the mean adjacent-period change, $\Delta Y_t = Y_t - Y_{t-1}$, with estimates $\widehat\pi_{g,t}$, $\widehat\mu_{g,t}$, and $\widehat\delta_{g,t}$ from parametric or semiparametric working models.

By Assumptions \ref{ass1} and \ref{ass3}, the untreated trend $\delta_{\infty,k}(X_k)$ is common to all groups not yet treated at period $k$, so all of them can serve as controls in period $k$. The simplest combination weights them in proportion to their generalized propensity scores, which ignores differences in the variability of their outcome changes; because parallel trends restrict only conditional means, less variable groups can be given more weight, as in inverse variance weighting in meta-analysis.

Let $\sigma^2_{g,t}(X_t) = \var(\Delta Y_t \mid X_t, G=g)$ be the variance of the outcome change in group $g$ and period $t$ conditional on time-varying covariates. Inspired by inverse variance weighting, we define weights for group-period cells
\[
    W_{l,k}(X_k) = \left[\sum_{s>k}\frac{\pi_{s,k}(X_k)}{\sigma^2_{s,k}(X_k)}\right]^{-1} \frac{\pi_{l,k}(X_k)}{\sigma^2_{l,k}(X_k)}.
\]
Let $\widehat{W}_{l,k}(X_k)$ be an estimate of $W_{l,k}(X_k)$ by plugging in the estimated $\pi_{\cdot,k}(X_k)$ and $\sigma^2_{\cdot,k}(X_k)$. We propose an augmented inverse variance weighting (AIVW) estimator of $\tau_{g,t}$ as
\begin{equation} \label{tauhat_gt}
\begin{aligned}
    \widehat\tau_{g,t} &= \frac{1}{\mathbb{P}_n\{I(G=g)\}} \mathbb{P}_n \bigg[ I(G=g) \sum_{k=g}^{t} \big\{ \Delta Y_k - \widehat\delta_{\infty,k}(X_k) \big\} \\
    & \qquad\qquad\qquad\qquad - \sum_{k=g}^{t} I(G>k)\frac{\widehat\pi_{g,k}(X_k)}{\widehat\pi_{G,k}(X_k)} \widehat{W}_{G,k}(X_k) \big\{\Delta Y_k - \widehat\delta_{\infty,k}(X_k) \big\}\bigg].
\end{aligned}
\end{equation}
The aggregated estimators $\widehat\tau_g^{\text{grp}}$, $\widehat\tau_t^{\text{prd}}$, $\widehat\tau_s^{\text{dyn}}$, and $\widehat\tau$ replace $\tau_{g,t}$ by $\widehat\tau_{g,t}$ and $P(G=g)$ by $\mathbb{P}_n\{I(G=g)\}$ in the definitions of Section \ref{sec2}.

If the variance is taken to be constant across groups and periods, $\sigma^2_{l,k}(X_k) = \sigma^2$ for all $l$ and $k$, the estimator reduces to an augmented inverse probability weighting (AIPW) estimator
\begin{equation}
\begin{aligned}
    \widehat\tau_{g,t} &= \frac{1}{\mathbb{P}_n\{I(G=g)\}} \mathbb{P}_n \bigg[ I(G=g) \sum_{k=g}^{t} \big\{ \Delta Y_k - \widehat\delta_{\infty,k}(X_k) \big\} \\
    & \qquad\qquad\qquad\qquad - \sum_{k=g}^{t} I(G>k) \frac{\widehat\pi_{g,k}(X_k)}{\sum_{l>k}\widehat\pi_{l,k}(X_k)} \big\{\Delta Y_k - \widehat\delta_{\infty,k}(X_k) \big\}\bigg].
\end{aligned}
\end{equation}
More generally, $W_{l,k}(X_k)$ depends on the conditional variances only through their ratios across not-yet-treated groups, so AIVW coincides with AIPW whenever $\sigma^2_{l,k}(X_k)$ is the same for all $l>k$, even if it varies with $X_k$ and $k$.

The construction differs from the doubly robust estimators of \citet{callaway2021difference} in three ways. First, the counterfactual trend in period $k$ is estimated from all groups with $G>k$. This allows the conditioning covariates to change from period to period but does not by itself reduce variance, because the adjacent-period changes of different control sets do not telescope into a single long difference. Second, the AIVW weights favor not-yet-treated groups with smaller conditional variance. Third, conditioning on $X_k$ in each period removes the component of the untreated trend explained by the time-varying covariates, which estimators that condition only on $X_0$ must treat as noise.

\subsection{Asymptotic properties of the proposed estimators}

We assume that $Y_t$ and $X_t$ have bounded total variation, so that the fitted models are bounded. Consistency requires that the summand of \eqref{tauhat_gt} belongs to a Glivenko--Cantelli class \citep{van2013weak}. Because it is Lipschitz in the nuisance functions, it suffices that the working models are Glivenko--Cantelli, as parametric models and many semiparametric and nonparametric ones, such as transformation models, kernel regression, and splines, are; these also satisfy the stronger Donsker condition \citep{lu2007estimation, athey2019generalized, kuchibhotla2020efficient, martinez2023efficient}.
Both AIVW and AIPW are doubly robust.

\begin{theorem}
Suppose the fitted models $\{\widehat\delta_{\cdot,\cdot}(\cdot), \widehat\pi_{\cdot,\cdot}(\cdot), \widehat\sigma^2_{\cdot,\cdot}(\cdot)\}$ and their true counterparts belong to a Glivenko--Cantelli class. Suppose that either the estimated conditional expected change model $\{\widehat\delta_{\cdot,\cdot}(\cdot)\}$ or the estimated propensity score model $\{\widehat\pi_{\cdot,\cdot}(\cdot)\}$ is $L_1$-consistent. Then $\widehat\tau_{g,t}$ is consistent for $\tau_{g,t}$, $\widehat\tau_g^{\text{grp}}$ is consistent for $\tau_g^{\text{grp}}$, $\widehat\tau_t^{\text{prd}}$ is consistent for $\tau_t^{\text{prd}}$, $\widehat\tau_s^{\text{dyn}}$ is consistent for $\tau_s^{\text{dyn}}$, and $\widehat\tau$ is consistent for $\tau$.
\end{theorem}

Consistency does not require a correctly specified variance model, because $\sigma^2_{\cdot,\cdot}(\cdot)$ enters only the weights. Let
\begin{align*}
    \phi_{g,t} &= \frac{1}{P(G=g)} \bigg[ I(G=g) \sum_{k=g}^{t} \big\{ \Delta Y_k - \delta_{\infty,k}(X_k) \big\} \\
    & \qquad\qquad\qquad - \sum_{k=g}^{t} I(G>k)\frac{\pi_{g,k}(X_k)}{\pi_{G,k}(X_k)} W_{G,k}(X_k) \big\{\Delta Y_k - \delta_{\infty,k}(X_k) \big\}\bigg].
\end{align*}
The influence functions of $\widehat\tau_{g,t}$, $\widehat\tau_g^{\text{grp}}$, $\widehat\tau_t^{\text{prd}}$, $\widehat\tau_s^{\text{dyn}}$, and $\widehat\tau$ are
\begin{align*}
    \varphi_{g,t} &= \phi_{g,t} - \frac{I(G=g)}{P(G=g)} \tau_{g,t}, \\
    \varphi_{g}^{\text{grp}} &= \frac{1}{T-g+1} \sum_{t=g}^{T} \bigg\{\phi_{g,t} - \frac{I(G=g)}{P(G=g)}\tau_g^{\text{grp}}\bigg\}, \\
    \varphi_{t}^{\text{prd}} &= \frac{1}{\sum_{g=1}^{t} P(G=g)} \sum_{g=1}^{t} \{P(G=g)\phi_{g,t} - I(G=g)\tau_t^{\text{prd}}\}, \\
    \varphi_s^{\text{dyn}} &= \frac{1}{\sum_{g=1}^{T-s+1} P(G=g)} \sum_{g=1}^{T-s+1} \{P(G=g)\phi_{g,g+s-1} - I(G=g)\tau_s^{\text{dyn}}\}, \\
    \varphi &= \frac{1}{\sum_{g=1}^{T} (T-g+1)P(G=g)} \sum_{g=1}^{T}\sum_{t=g}^{T} \{P(G=g)\phi_{g,t} - I(G=g)\tau\},
\end{align*}
respectively. These influence functions have zero means. The asymptotic distribution of the proposed estimator is given in the following theorem.

\begin{theorem}
Suppose that the fitted models $\{\widehat\delta_{\cdot,\cdot}(\cdot), \widehat\pi_{\cdot,\cdot}(\cdot), \widehat\sigma^2_{\cdot,\cdot}(\cdot)\}$ and their true counterparts belong to a Donsker class. Suppose these fitted models converge to the true value at a rate of $o_p(n^{-1/4})$ in $L_2$-norm. Then
$\sqrt{n} (\widehat\tau_{g,t} - \tau_{g,t}) \xrightarrow{d} N(0, \mathbb{P}\varphi_{g,t}^2)$,
$\sqrt{n} (\widehat\tau_g^{\text{grp}} - \tau_g^{\text{grp}}) \xrightarrow{d} N(0, \mathbb{P}{\varphi_{g}^{\text{grp}}}^2)$,
$\sqrt{n} (\widehat\tau_t^{\text{prd}} - \tau_t^{\text{prd}}) \xrightarrow{d} N(0, \mathbb{P}{\varphi_{t}^{\text{prd}}}^2)$, 
$\sqrt{n} (\widehat\tau_s^{\text{dyn}} - \tau_s^{\text{dyn}}) \xrightarrow{d} N(0, \mathbb{P}{\varphi_{s}^{\text{dyn}}}^2)$, and
$\sqrt{n} (\widehat\tau - \tau) \xrightarrow{d} N(0, \mathbb{P}\varphi^2)$.
\end{theorem}

For machine learning working models, the Donsker condition can be avoided by sample splitting and cross-fitting \citep{chernozhukov2018double, chernozhukov2022locally}. The proofs of Theorems 2 and 3 are given in Supplementary Material A.2 and A.3. The variance of each estimator is estimated by $n^{-1}\mathbb{P}_n\widehat\varphi^2$, where $\widehat\varphi$ is the corresponding fitted influence function.

The following two remarks give sufficient conditions under which AIVW attains the semiparametric efficiency bound.

\begin{remark} \label{rmk1}
Suppose that, in place of Assumptions \ref{ass2}--\ref{ass3}, parallel trends hold conditional on the complete observed history $H_t = (X_0, Y_0, \ldots, X_{t-1}, Y_{t-1}, X_t)$, without requiring $H_t(g)=H_t(\infty)$:
\[
E\{\Delta Y_t(\infty) \mid H_t, G\} = E\{\Delta Y_t(\infty) \mid H_t, G=\infty\}.
\]
This assumption neither implies nor is implied by Assumptions \ref{ass2}--\ref{ass3}; because $H_t$ contains outcomes realized after treatment begins, it restricts the untreated change given a treatment-affected history \citep{lechner2011estimation}. The analog of Theorem 1 with $X_k$ replaced by $H_k$ still identifies the ATT (Supplementary Material A.4), and the AIVW estimator becomes
\begin{equation}
\begin{aligned}
    \widehat\tau_{g,t} &= \frac{1}{\mathbb{P}_n\{I(G=g)\}} \mathbb{P}_n \bigg[ I(G=g) \sum_{k=g}^{t} \big\{ \Delta Y_k - \widehat\delta_{\infty,k}(H_k) \big\} \\
    & \qquad\qquad\qquad\qquad - \sum_{k=g}^{t} I(G>k)\frac{\widehat\pi_{g,k}(H_k)}{\widehat\pi_{G,k}(H_k)} \widehat{W}_{G,k}(H_k) \big\{\Delta Y_k - \widehat\delta_{\infty,k}(H_k) \big\}\bigg], 
    \end{aligned}
\end{equation}
where $\delta_{g,t}$, $\pi_{g,t}$, and $\sigma^2_{g,t}$ are fitted as functions of $H_t$. The likelihood of $O = (G,H_T,Y_T)$ factorizes as
\[
f(G,H_T,Y_T) = f(G,H_0) f(Y_0 \mid G, H_0) \prod_{t=1}^{T} f(H_t \mid G, H_{t-1}, Y_{t-1}) f(Y_t \mid G, H_t).
\]
The influence function $\varphi_{g,t}$ obtained by replacing $X_k$ with $H_k$ lies in the tangent space of this model, with a moment restriction on $f(Y_t \mid G, H_t)$ for $G>t$, provided that for $1 \leq s \leq g-1$ the conditional mean $E[\sum_{k=g}^{t}\{\delta_{g,k}(H_k)-\delta_{\infty,k}(H_k)\} \mid G=g, H_s, \Delta Y_s]$ does not depend on $\Delta Y_s$, that is, the history-conditional effect in group $g$ is not predicted by a pre-treatment outcome change beyond the history preceding it. Under this condition, $\varphi_{g,t}$ is the efficient influence function (EIF) of $\tau_{g,t}$ \citep{tsiatis2006semiparametric, van2000asymptotic}, and $\widehat\tau_{g,t}$ attains the semiparametric efficiency bound in the model restricted only by history-conditional parallel trends. If the condition holds for every cell entering an aggregate (it is vacuous for $g=1$), the aggregated estimators are also efficient. Supplementary Material A.4 gives the necessary and sufficient covariance condition, the EIF when it fails, and the proofs.
\end{remark}

\begin{remark} \label{rmk2}
Under Assumption \ref{ass3}, which restricts only a conditional first moment, the EIF has not been derived in general, and we make no general efficiency claim; \citet{chen2025efficient} derive it when parallel trends are conditional on baseline covariates. Suppose, in addition, that given $(G,X_t)$, the change $\Delta Y_t$ is generated independently of the past and the covariates evolve independently of past outcomes, so that
\begin{align*}
f(G,H_T,Y_T) &= f(G,X_0) f(Y_0 \mid G,X_0) \prod_{t=1}^{T} f(X_t \mid G,\bar{X}_{t-1}) f(\Delta Y_t \mid G,X_t),
\end{align*}
where $\bar{X}_{t-1}=(X_0,\ldots,X_{t-1})$. Then $\varphi_{g,t}$ lies in the tangent space of this model with a moment restriction on $f(\Delta Y_t \mid G, X_t)$ for $G>t$, so it is the EIF of $\tau_{g,t}$ (Supplementary Material A.4), and $\widehat\tau_{g,t}$ and the aggregated estimators attain the semiparametric efficiency bound in the model restricted by parallel trends and these two independence conditions.
\end{remark}

Under misspecification, suppose $\widehat\delta_{\cdot,\cdot}(\cdot)$, $\widehat\pi_{\cdot,\cdot}(\cdot)$, and $\widehat\sigma^2_{\cdot,\cdot}(\cdot)$ converge to $\bar\delta_{\cdot,\cdot}(\cdot)$, $\bar\pi_{\cdot,\cdot}(\cdot)$, and $\bar\sigma^2_{\cdot,\cdot}(\cdot)$, respectively, in $L_2$-norm with influence functions $\varphi_{\delta}$, $\varphi_{\pi}$, and $\varphi_{\sigma^2}$, where at least one of $\bar\delta_{\cdot,\cdot}(\cdot)$ and $\bar\pi_{\cdot,\cdot}(\cdot)$ is the true value. We modify the influence functions $\overline\varphi_{g,t}$, $\overline\varphi_{g}^{\text{grp}}$, $\overline\varphi_{t}^{\text{prd}}$, $\overline\varphi_{s}^{\text{dyn}}$, and $\overline\varphi$ by replacing true models with their limits. The derivatives of the influence functions with respect to the nuisance models are mean-zero at the true values but not at misspecified limits; to account for this first-order term, let
\begin{align*}
\dot\phi_{g,t} &= \mathbb{P} \nabla \phi_{g,t}\{\bar\delta_{\cdot,\cdot}(\cdot), \bar\pi_{\cdot,\cdot}(\cdot), \bar\sigma^2_{\cdot,\cdot}(\cdot)\}[\varphi_{\delta}, \varphi_{\pi}, \varphi_{\sigma^2}],
\end{align*}
be the Hadamard derivative of $\phi_{g,t}$ with respect to nuisance models at their limits along the paths of their influence functions; see Supplementary Material A.5 for details. To ensure the existence of influence functions, fitted models that converge at rates slower than $O_p(n^{-1/2})$ are not allowed. If all working models are parametric, then $\dot\phi_{g,t}$ is a linear combination of the influence functions of the model parameters. We define $\dot\phi_{g}^{\text{grp}}$, $\dot\phi_{t}^{\text{prd}}$, $\dot\phi_{s}^{\text{dyn}}$, and $\dot\phi$ similarly based on the cell proportions. The following result gives the asymptotic distributions of the ATT estimators.

\begin{theorem}
Suppose that the fitted models $\{\widehat\delta_{\cdot,\cdot}(\cdot), \widehat\pi_{\cdot,\cdot}(\cdot), \widehat\sigma^2_{\cdot,\cdot}(\cdot)\}$ and their limiting counterparts $\{\bar\delta_{\cdot,\cdot}(\cdot), \bar\pi_{\cdot,\cdot}(\cdot), \bar\sigma^2_{\cdot,\cdot}(\cdot)\}$ belong to a Donsker class. Suppose these fitted models converge in $L_2$-norm with influence functions $\{\varphi_{\delta}, \varphi_{\pi}, \varphi_{\sigma^2}\}$. Then
$\sqrt{n} (\widehat\tau_{g,t} - \tau_{g,t}) \xrightarrow{d} N(0, \mathbb{P}\{\overline\varphi_{g,t}+\dot\phi_{g,t}\}^2)$,
$\sqrt{n} (\widehat\tau_g^{\text{grp}} - \tau_g^{\text{grp}}) \xrightarrow{d} N(0, \mathbb{P}\{\overline\varphi_{g}^{\text{grp}}+\dot\phi_{g}^{\text{grp}}\}^2)$,
$\sqrt{n} (\widehat\tau_t^{\text{prd}} - \tau_t^{\text{prd}}) \xrightarrow{d} N(0, \mathbb{P}\{\overline\varphi_{t}^{\text{prd}}+\dot\phi_{t}^{\text{prd}}\}^2)$, 
$\sqrt{n} (\widehat\tau_s^{\text{dyn}} - \tau_s^{\text{dyn}}) \xrightarrow{d} N(0, \mathbb{P}\{\overline\varphi_{s}^{\text{dyn}}+\dot\phi_{s}^{\text{dyn}}\}^2)$, and
$\sqrt{n} (\widehat\tau - \tau) \xrightarrow{d} N(0, \mathbb{P}\{\overline\varphi+\dot\phi\}^2)$.
\end{theorem}

Because the asymptotic variances involve the true nuisance functions, they cannot be estimated consistently under misspecification; Theorem 4 guarantees asymptotic normality, so the bootstrap can be used for inference.

\subsection{The general covariance-weighted class}

The weights $W_{l,k}(X_k)$ are the generalized least squares weights for combining the not-yet-treated groups within period $k$. Each group $l>k$ supplies a moment for the common trend,
\[
m_{l,k} = \frac{I(G=l)}{\pi_{l,k}(X_k)}\{\Delta Y_k - \delta_{\infty,k}(X_k)\}, \qquad E(m_{l,k} \mid X_k) = 0,
\]
where the conditional mean is zero by Assumptions \ref{ass1} and \ref{ass3}. Let $m_k = (m_{l,k})_{l>k}$ and $\Sigma_k(X_k) = \cov(m_k \mid X_k)$. Among linear combinations $w_k^{\top}m_k$ with weights that sum to one, the conditional variance is minimized by the generalized least squares weights
\begin{equation}
w_k^*(X_k) \;=\; \frac{\Sigma_k(X_k)^{-1}\mathbf{1}}{\mathbf{1}^\top \Sigma_k(X_k)^{-1}\mathbf{1}},
\label{eq:gls-weights}
\end{equation}
where $\mathbf{1}$ is a vector of ones. Because each unit belongs to exactly one group, $\Sigma_k(X_k)$ is diagonal with entries $\sigma^2_{l,k}(X_k)/\pi_{l,k}(X_k)$, and the $l$-th coordinate of $w_k^*(X_k)$ equals $W_{l,k}(X_k)$; indeed, the control term of \eqref{tauhat_gt} for period $k$ equals $\pi_{g,k}(X_k)\sum_{l>k}W_{l,k}(X_k)m_{l,k}$. The weights ignore the dependence of outcome changes across periods and the information in the pre-treatment changes of the treated group (Supplementary Material A.4), so AIVW is efficient only under the conditions of Remarks \ref{rmk1} and \ref{rmk2} and need not improve on AIPW when outcome changes are serially correlated (Section \ref{sec4}).

The covariance-weighted estimator of \citet{chen2025efficient} optimizes over a larger set of comparisons: under parallel trends conditional on baseline covariates, every pre-treatment period can serve as a baseline for $\tau_{g,t}$, and their weights exploit the correlation of these comparisons. When the time-varying covariates carry no further information about the untreated trend, their estimator can therefore be more precise than AIVW. The comparison reverses when parallel trends require conditioning on $X_k$, so that baseline-adjusted comparisons are invalid, or when $X_k$ predicts the untreated trend, so that conditioning on it removes variation that no weighting of baseline-adjusted comparisons can remove (Section \ref{sec4}). Replacing $W_{l,k}(X_k)$ by any weights that sum to one over $l>k$ preserves double robustness and asymptotic normality.

\subsection{Connection with linear outcome regression models} \label{sec:linear}

To make the outcome regression more flexible, the treatment indicator term in the two-way fixed effects model can be split into several dummy variables representing the length of exposure, thereby capturing heterogeneous effects across exposure lengths \citep{sun2021estimating}. Furthermore, we add the interaction terms between covariates and treatment as well as between covariates and periods \citep{wooldridge2025two}, so the working model for outcome regression is
\begin{equation*}
Y_{t} = \alpha D_{t} + \sum_{k=1}^{T-1} \alpha_k I(t-G=k) + \lambda_{t} + \gamma_{G} + \beta_1^{\top} X_t + \beta_2^{\top} X_tD_t + \beta_3^{\top} X_tt + \varepsilon_{t}.
\end{equation*}
Note that interactions between groups and periods should not be included in the model to respect parallel trends in observed outcomes for untreated units. The coefficient $\alpha$ is no longer the treatment effect. Under this working model, the conditional mean outcome in group $g$ and period $t$ is
\[
\mu_{g,t}(X_t) = \alpha I(g\leq t) + \sum_{k=1}^{T-1} \alpha_k I(t-g=k) + \lambda_{t} + \gamma_{g} + \beta_1^{\top} X_t + \beta_2^{\top} X_tI(g\leq t) + \beta_3^{\top} X_tt,
\]
and the counterfactual mean outcome under control is
\[
\mu_{g,t}^0(X_t) = \lambda_t + \gamma_g + \beta_1^{\top} X_t + \beta_3^{\top} X_tt.
\]
We approximate the conditional mean change in outcomes $\delta_{\infty,k}(X_k)$ that appears in the influence functions by $\mu^0_{g,k}(X_k) - \mu^0_{g,k-1}(X_{k-1})$, which does not rely on $g$, in line with the parallel trends assumption. This is a function of $X_k$ when the components of $X_{k-1}$ that enter $\mu^0_{g,k-1}$ are contained in $X_k$; otherwise, as when lagged covariates enter the outcome regression, Assumption \ref{ass3} is applied with $X_k$ augmented by those components. If there is no missingness, the sample for outcome regression includes $n(T+1)$ observations indexed by $\{(i,t): i=1,\ldots,n, t=0,\ldots,T\}$, which we pretend to be independent when fitting the model. Let $\widehat\mu_{g,t}^0(X_t)$ be the fitted counterfactual mean outcome under control in group $g$ and period $t$. 

The AIVW estimator can be obtained by weighting the residuals from the outcome regression. For the group-period ATT, we define 
\begin{align*}
H_{G,t}^{g,r} &= I(G=g)\{I(t=r)-I(t=g-1)\} \\
&\quad - \sum_{k=g}^{r}I(G>k)\frac{\pi_{g,k}(X_k)}{\pi_{G,k}(X_k)}W_{G,k}(X_k)\{I(t=k)-I(t=k-1)\}.
\end{align*}
Then the influence function of $\tau_{g,r}$ can be written as
\begin{equation}
\varphi_{g,r} = \frac{1}{P(G=g)}\sum_{t=0}^{T} H_{G,t}^{g,r} \{Y_t-\mu_{G,t}^0(X_t)-I(G=g,t=r)\tau_{g,r}\}.
\end{equation}

Let $\widehat\varepsilon_t$ be the residual of the outcome regression. To estimate the weights, we fit
\[
\log\{(\widehat\varepsilon_t-\widehat\varepsilon_{t-1})^2\} = \lambda_t^* + \gamma_G^* + \beta^{*\top} X_t + \epsilon_t
\]
among untreated observations $\{(i,t): D_{it}=0, t>0\}$ and set $\widehat\sigma_{g,t}^2(X_t) = \exp(\widehat\lambda_t^* + \widehat\gamma_g^* + \widehat\beta^{*\top} X_t)$. Only the group effects $\gamma_g^*$ enter the weights, because $\lambda_t^*$ and $\beta^{*\top}X_t$ are common to all groups and cancel from $W_{G,t}(X_t)$. With equal variances across groups, AIVW reduces to AIPW, which fits fewer models and avoids the finite-sample variability of an estimated variance model; an $F$-test of equal group effects $\gamma_g^*$ can be used to decide whether AIVW is needed.

We fit the time-varying propensity score $\pi_{g,t}(X_t)$ by ordinal logistic regression,
\[
\frac{P(G \leq k \mid X_t)}{P(G > k \mid X_t)} = \frac{\sum_{s\leq k}\pi_{s,t}(X_t)}{\sum_{s>k}\pi_{s,t}(X_t)} = \exp(\zeta_{kt} + \zeta_{t}^{\top}X_t),
\]
denoted by $\widehat\pi_{g,t}(X_t)$; other ordinal models, such as the adjacent-category logit used in our simulations, can be used in the same way. Because these working models are parametric, the conditions of Theorem 3 hold under correct specification. With the fitted models plugged into $H_{G,t}^{g,r}$, $\widehat\tau_{g,r}$ solves $\mathbb{P}_n\widehat\varphi_{g,r} = 0$: it is the sum of $\widehat{H}_{G_i,t}^{g,r}\{Y_{it}-\widehat\mu_{G_i,t}^0(X_{it})\}$ over all units and periods, divided by the number of units with $G_i=g$.

For the overall ATT, we define
\[
H_{G,t} = \sum_{r=1}^{T}\sum_{g=1}^{r} H_{G,t}^{g,r}.
\]
Under the homoskedastic working model underlying AIPW,
\begin{align*}
H_{G,t} &= D_t - (T-G+1)I(t=G-1) \\
&\quad - \frac{\sum_{s\leq t}\pi_{s,t}(X_t)}{\sum_{s>t}\pi_{s,t}(X_t)}(T-t+1)(1-D_t) + \frac{\sum_{s\leq t+1}\pi_{s,t+1}(X_{t+1})}{\sum_{s>t+1}\pi_{s,t+1}(X_{t+1})}(T-t)(1-D_{t+1}).
\end{align*}
The last term is zero for $t=T$. The function $H_{G,t}$ involves the propensity score only through the cumulative odds $\sum_{s\leq t}\pi_{s,t}(X_t)/\sum_{s>t}\pi_{s,t}(X_t)$, which the ordinal logistic model specifies directly. Then the influence function of $\tau$ can be written as
\begin{equation} \label{IF_tau_H}
\varphi = \frac{1}{(T+1)P(D_t=1)}\sum_{t=0}^{T} H_{G,t}\{Y_t-\mu_{G,t}^0(X_t)-D_t\tau\}.
\end{equation}
Here $P(D_t=1) = \sum_{g=1}^{T}(T-g+1)P(G=g)/(T+1)$ is the probability of being treated for all $(i,t)$ pairs. Accordingly, $\widehat\tau$ solves $\mathbb{P}_n\widehat\varphi = 0$: it is the sum of $\widehat{H}_{G_i,t}\{Y_{it}-\widehat\mu_{G_i,t}^0(X_{it})\}$ over all units and periods, divided by the number of treated unit-periods.

\section{Simulation Studies} \label{sec4}

We compare AIPW and AIVW with five estimators: TWFE with time-varying covariates; the doubly robust estimators of \citet{callaway2021difference} with never-treated (DRnt) and not-yet-treated (DRny) controls; the estimator of \citet{chen2025efficient} (CSX), which is efficient under parallel trends conditional on baseline covariates; and a long-difference doubly robust estimator (LD) in the spirit of \citet{caetano2022difference} and \citet{caetano2024difference}, which conditions $Y_t-Y_{g-1}$ on the time-invariant covariates and on the time-varying covariate at its two endpoints. As proposed by their authors, DRnt, DRny, and CSX condition on covariates measured in the baseline period.

\subsection{Designs}

All four designs have five periods, $t=0,\ldots,4$, adoption groups $G\in\{1,\ldots,4\}$, and never-treated units, coded $G=5$. Adoption depends on two time-invariant covariates $Z_1$ and $Z_2$, and a time-varying covariate $Z_{3,t}$ follows a random walk with a group-specific drift $\lambda_{g,t}$ that does not respond to treatment, so Assumption \ref{ass2} holds; the treatment effect grows with the length of exposure and varies with $Z_{3,t}$. 

In Designs 1 and 2, $Z_{3,t}$ shifts the level of the untreated outcome. The drift is $\lambda_{g,t}=0.1(3-g)$ in Design 1 and zero in Design 2, so parallel trends conditional on baseline covariates fail in Design 1 but hold in Design 2. In Designs 3 and 4, $Z_{3,t}$ shifts the untreated change, and the long difference depends on the path of $Z_{3,t}$ between its endpoints. The drift $\lambda_{g,t}=0.1(3-g)$ is constant in Design 3, so that the long-difference assumption holds; the drift $\lambda_{g,t}=0.3(3-g)(-1)^{t+1}$ changes sign in Design 4, so that the long-difference assumption fails.

Assumption \ref{ass3} holds in all four designs. Therefore, Design 1 concerns consistency relative to estimators that adjust for baseline covariates. Design 2 concerns estimation precision when all doubly robust estimators are consistent. Design 3 concerns estimation precision relative to LD when both the period-by-period and the long-difference assumptions hold. Design 4 concerns consistency when only the period-by-period assumption holds. 

Besides a unit random effect, the outcome contains an error $u_t$ with one of three structures: (i) independent $N(0,1)$ errors; (ii) heteroskedastic errors, independent across periods, whose standard deviation decreases with $G$ and varies with $Z_2$ and $Z_{3,t}$; and (iii) cumulative errors, $u_t=\sum_{k\leq t}u^*_k$ with independent increments whose standard deviation decreases with $G$ at the same rate as in (ii). We cross each design with the three error structures and $n\in\{100,500\}$, with 1,000 replications per setting, and apply all estimators to the same data sets. 

Supplementary Material C.1 and C.2 give the data-generating processes and the implementation of the estimators. Supplementary Material C.3 reports the bias, standard deviation, average standard error, and coverage of the overall ATT in each setting.

\subsection{Results}

\begin{figure}[!tbhp]
\centering
\includegraphics[width=\textwidth]{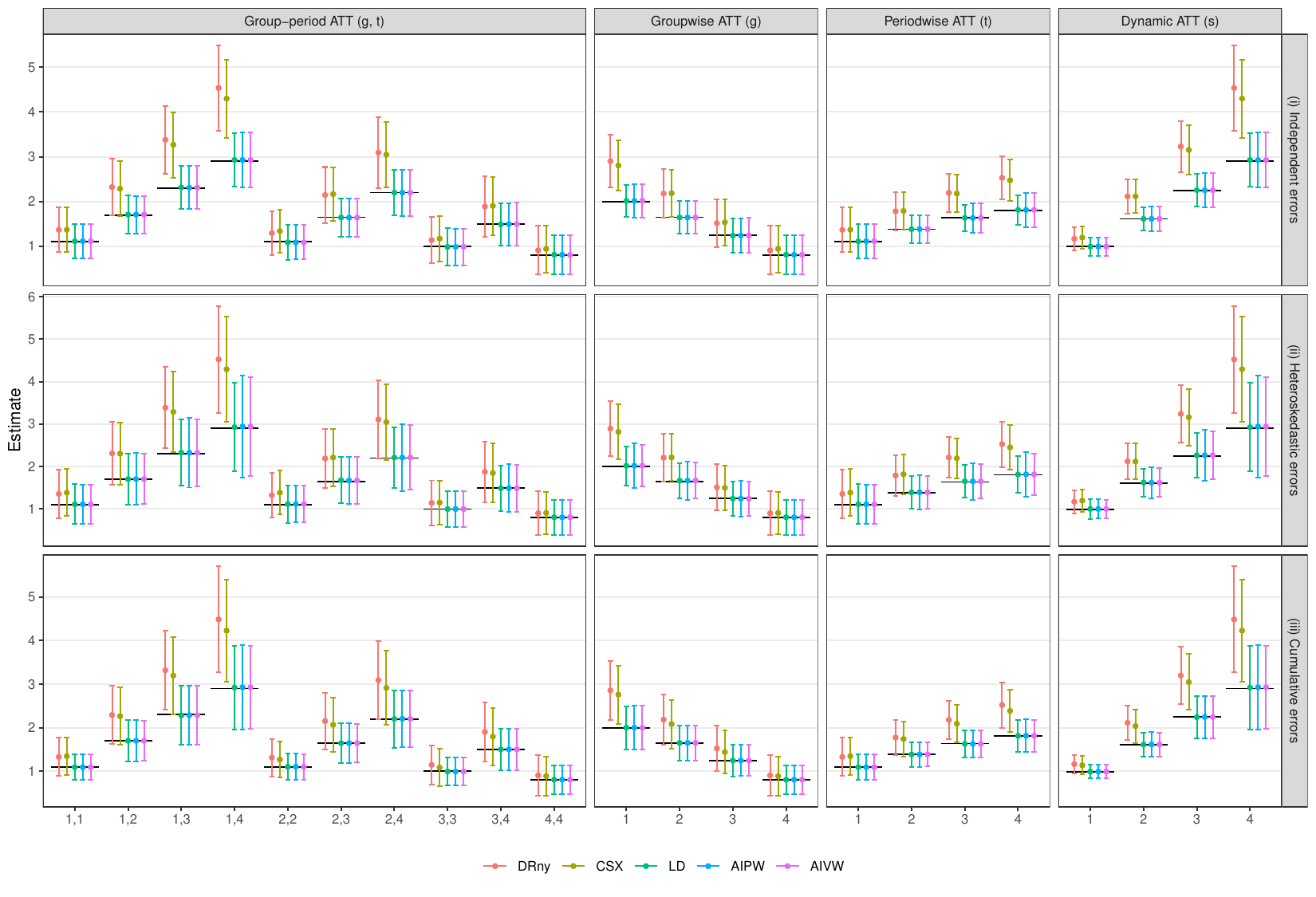}
\caption{Group-period, groupwise, periodwise, and dynamic ATTs in Design 1 for each error structure (rows), with $n=500$. Points are average estimates, and error bars are average estimates $\pm 1.96$ average estimated standard errors. Black horizontal lines are the true values.}
\label{fig:sim-cells1}
\end{figure}

\begin{figure}[!tbhp]
\centering
\includegraphics[width=\textwidth]{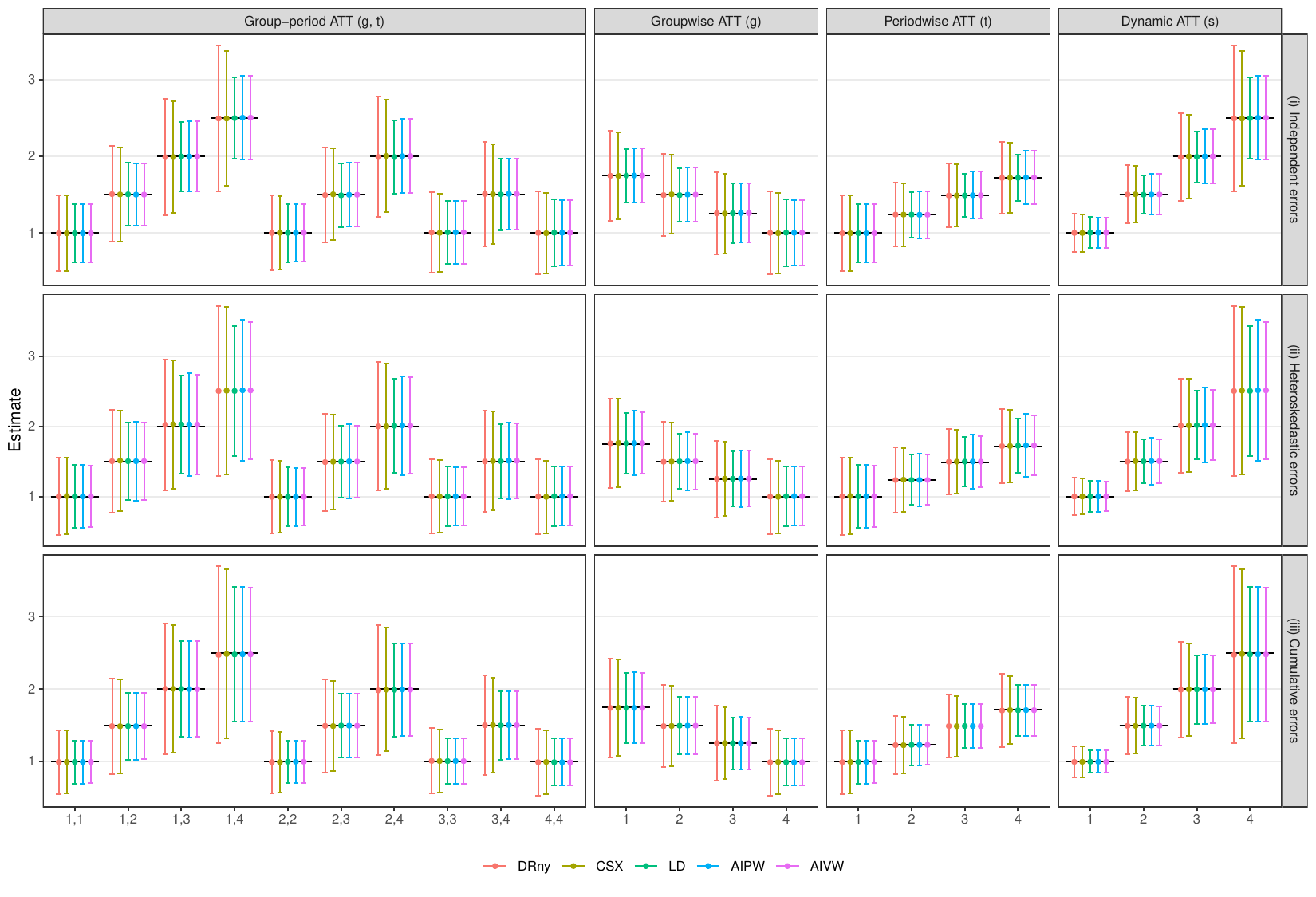}
\caption{Group-period, groupwise, periodwise, and dynamic ATTs in Design 2 for each error structure (rows), with $n=500$. Points are average estimates, and error bars are average estimates $\pm 1.96$ average estimated standard errors. Black horizontal lines are the true values.}
\label{fig:sim-cells2}
\end{figure}

\begin{figure}[!tbhp]
\centering
\includegraphics[width=\textwidth]{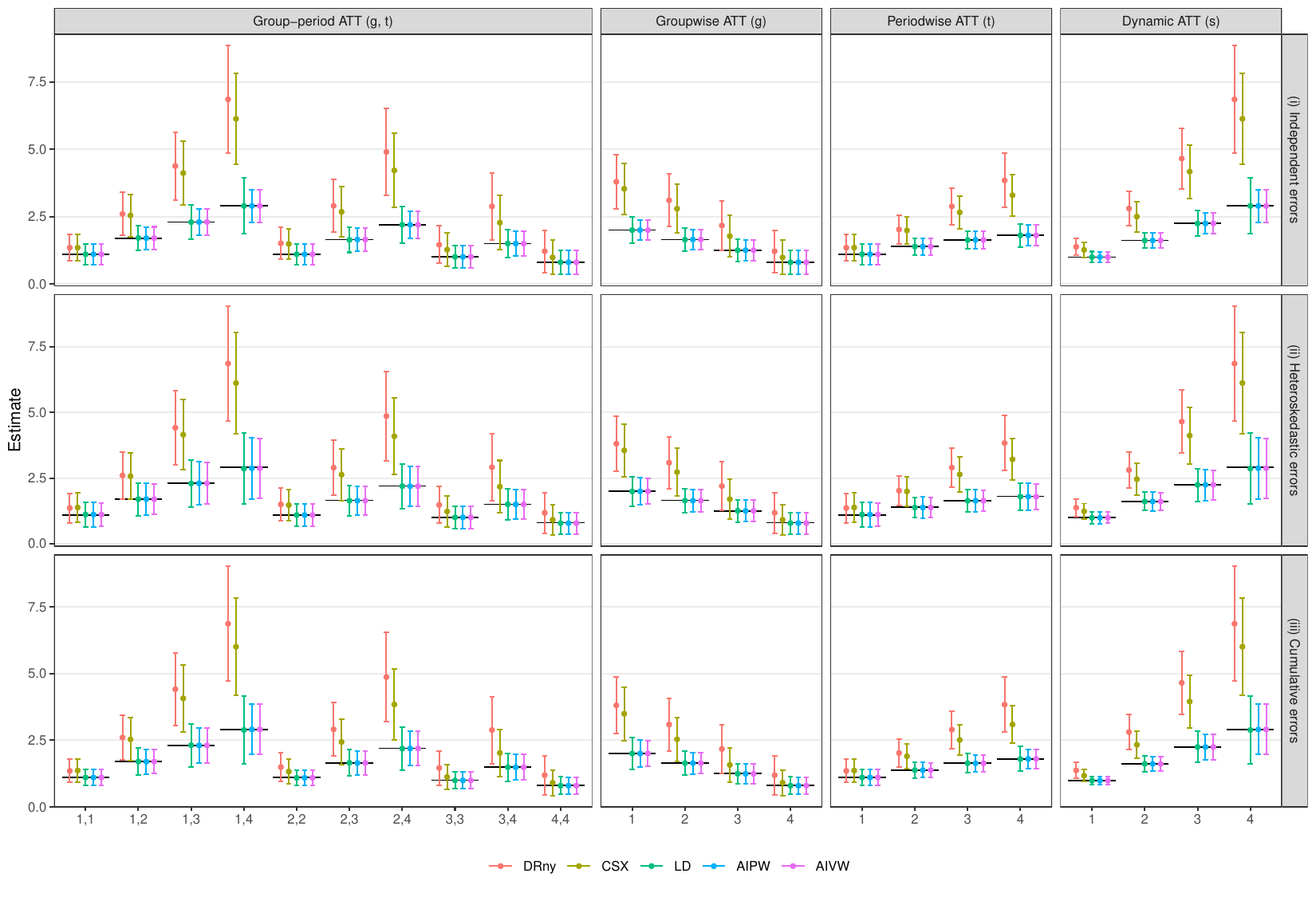}
\caption{Group-period, groupwise, periodwise, and dynamic ATTs in Design 3 for each error structure (rows), with $n=500$. Points are average estimates, and error bars are average estimates $\pm 1.96$ average estimated standard errors. Black horizontal lines are the true values.}
\label{fig:sim-cells3}
\end{figure}

\begin{figure}[!tbhp]
\centering
\includegraphics[width=\textwidth]{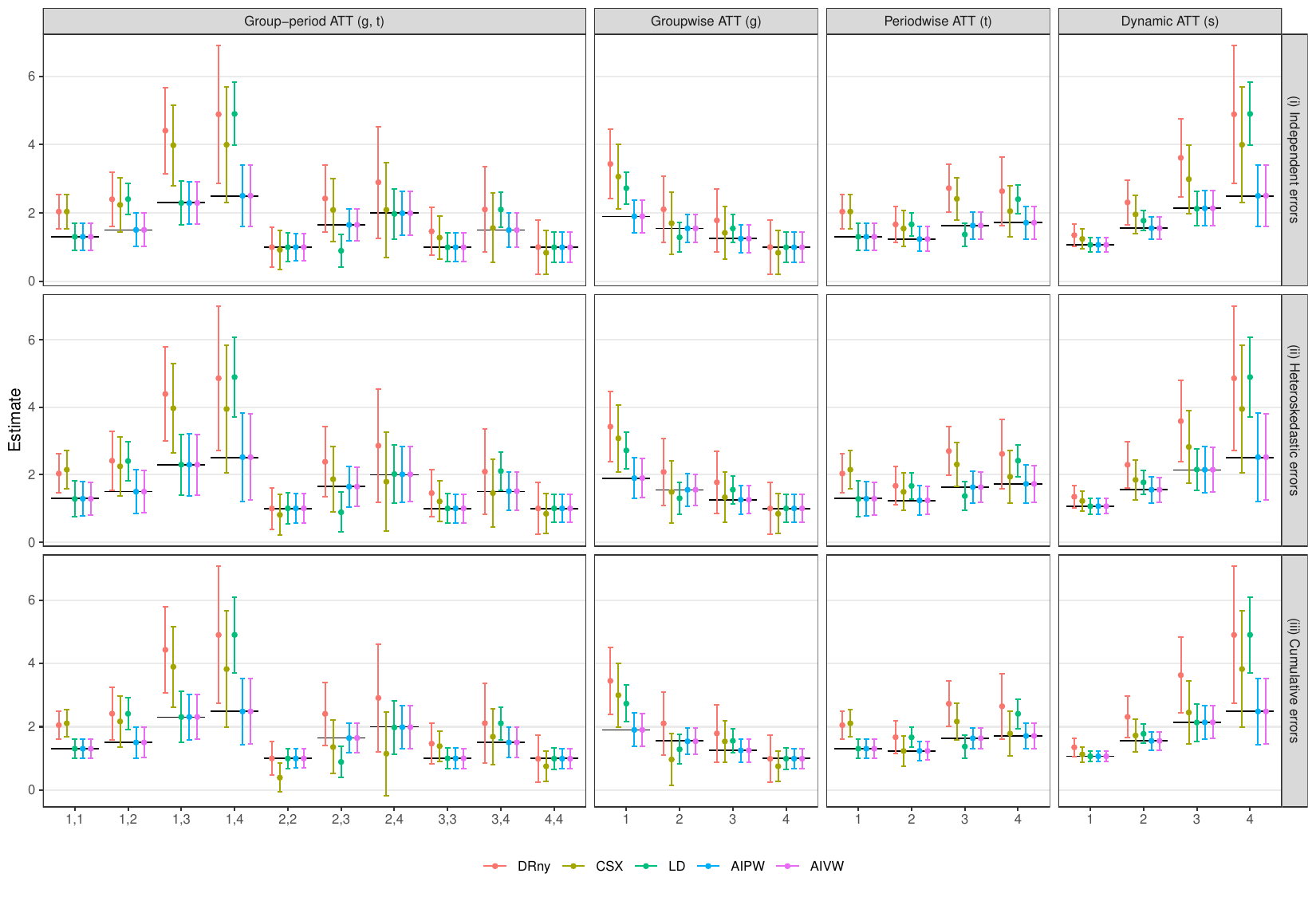}
\caption{Group-period, groupwise, periodwise, and dynamic ATTs in Design 4 for each error structure (rows), with $n=500$. Points are average estimates, and error bars are average estimates $\pm 1.96$ average estimated standard errors. Black horizontal lines are the true values.}
\label{fig:sim-cells4}
\end{figure}

Figures \ref{fig:sim-cells1}--\ref{fig:sim-cells4} report the group-period, groupwise, periodwise, and dynamic ATTs of DRny, CSX, LD, AIPW, and AIVW with $n=500$. We do not report DRnt and TWFE in these figures because DRnt is less precise than DRny, and TWFE has no cell-level estimates. Tables S4--S7 in the Supplementary Material cover both sample sizes.

\emph{Design 1.} TWFE, DRnt, DRny, and CSX are biased. Efficient weighting in CSX does not remove bias from the conditioning set. AIPW, AIVW, and LD, which condition on the time-varying covariate, are asymptotically unbiased.

\emph{Design 2.} All doubly robust estimators are asymptotically unbiased. The variances of DRny and CSX are 1.47--2.95 times that of AIPW for the overall ATT, and 1.37--3.57 times for the cell-level targets. AIVW reduces the variance of AIPW by 16--17\% under error structure (ii) and changes it by at most 2\% under (i) and (iii). AIVW shifts weight toward later-adopting groups because the error variance decreases with $G$. Of note, pooling not-yet-treated groups period by period does not by itself reduce variance. LD, which absorbs $Z_{3,t}-Z_{3,g-1}$ by conditioning on the endpoint covariates, has 0.68--1.00 times the variance of AIPW for the overall ATT and 0.77--1.08 times for the cell-level targets.

\emph{Design 3.} TWFE, DRnt, DRny, and CSX are biased, whereas LD, AIPW, and AIVW are asymptotically unbiased. For the overall ATT, the variance of LD is 1.11--1.40 times that of AIPW under error structures (i) and (iii), because the long difference retains the variation of the intermediate covariate values, which the period-by-period construction in AIPW and AIVW removes. For the cell-level targets, the variance of LD is 0.94--3.64 times that of AIPW. AIVW reduces the variance of AIPW under error structure (ii).

\emph{Design 4.} DRnt, DRny, LD, and CSX are biased, and the bias becomes larger with the length of exposure. AIPW and AIVW are asymptotically unbiased. AIVW reduces the variance of AIPW under error structure (ii).

\section{Application to NCEE (Gaokao) data} \label{sec5}

Many countries assign students to schools and universities through centralized matching mechanisms. Under the Immediate Acceptance (IA), or Boston mechanism, schools immediately and irrevocably accept their highest-priority applicants in each round \citep{abdulkadirouglu2003school}. In recent decades, Chinese provinces have moved from IA to a parallel mechanism, a variant of Deferred Acceptance in which students list several universities within each choice band and assignments within a band are tentative \citep{chen2020empirical, ha2020college}. Previous research suggests that the parallel mechanism improves stability and reduces manipulability \citep{kang2020matching}, but these studies rely on two-way fixed effects models or event studies, which can be biased under heterogeneous treatment effects and do not let parallel trends depend on time-varying features of the admission process, such as the number of students in a province and track.

We use administrative data on the National College Entrance Examination (NCEE, or \textit{Gaokao}) from 2007 to 2011 for the 27 provinces that used IA or the parallel mechanism. All used IA in 2007; 3, 10, 6, and 2 provinces adopted the parallel mechanism in 2008, 2009, 2010, and 2011, respectively, and six had not adopted it by 2011. Splitting each province into its STEM and non-STEM tracks gives $n=54$ units; we call each province-track-year a game.

We use justified envy (JE) to quantify fairness. JE is a standard measure of fairness in the school choice literature \citep{balinski1999tale, abdulkadirouglu2003school, kamada2024fair}. We say that student $i$ justifiably envies student $j$ for school $s$ if $i$ would rather be assigned to school $s$, where some student $j$, who has a lower priority (i.e., lower score) than $i$, is assigned. In this case, student $i$ is a blocking student; student $i$ and school $s$ are a blocking pair; student $i$, $j$, and school $s$ are a blocking triplet. We consider four numerical outcomes for justified envy: (1) BS, the total number of blocking students in a game; (2) BP, the total number of blocking pairs in a game; (3) TE, the total tridimensional envy of blocking triplets in a game, where the tridimensional envy is defined as the diagonal distance of student quantiles and university tiers in a blocking triplet; (4) BT, the total number of blocking triplets in a game. Formal definitions of these measures appear in \citet{kang2020matching}.

The baseline covariates are province-level log GDP per capita and population (2007 values from the \textit{China Statistical Yearbook 2008}), measured before any adoption, and the track. The time-varying covariates are two policy-timing indicators: whether preferences are submitted after the exam and after scores are known, and the number of students in the game. Because treatment is assigned by province, the propensity score depends only on log GDP per capita and population. The outcome regressions in AIPW, AIVW, and the TWFE model include all covariates as well as lagged time-varying covariates. DRnt, DRny, and CSX adjust for baseline covariates and the number of students before treatment. LD regresses $Y_t-Y_{g-1}$ on the baseline covariates and the number of students in periods $g-1$ and $t$. Standard errors are clustered by province (27 clusters). 

Assumption \ref{ass2} requires that these covariates do not respond to the admission mechanism. Provincial education authorities set the timing indicators as part of the admissions calendar, a decision separate from the choice of mechanism, which does not prescribe when preferences are submitted. Seven provinces changed their timing rule during the sample period, all to submission after scores are known; five did so in a year other than their adoption year, but Henan and Shaanxi did so in 2010, the year they adopted the parallel mechanism, so for them a joint policy decision cannot be ruled out. The number of students is determined mainly by the size of the birth cohort and by track quotas set years in advance, whereas the mechanism changes only how students who sit the examination are matched to universities. Neither argument can be tested with the data.

The $F$-test of Section \ref{sec:linear} does not reject equal variances across groups for any outcome ($P$ between 0.15 and 0.81). The residuals of the outcome regression in untreated observations show no group-specific linear trends ($P$ between 0.25 and 0.99; Supplementary Material B.1), which is consistent with, but cannot prove, parallel trends for the post-treatment counterfactuals.

Table \ref{tab:data1} reports the overall ATTs. Because the $F$-test does not reject, we treat AIPW as the primary estimator. AIPW estimates that the parallel mechanism reduces BP, TE, and BT relative to IA ($P \leq 0.001$), and all other estimators agree in sign and significance. For BS, the AIPW estimate is negative but fragile, at $-0.605$ ($P = 0.045$), whereas LD is $-0.743$ ($P = 0.024$). The estimates of AIVW, DRnt, DRny, and CSX are not significant ($P \geq 0.066$). The number of students is the time-varying covariate that matters most: its changes explain 83\% and 81\% of the variance of untreated changes in BP and TE beyond period effects and baseline covariates.

\begin{table}
\centering
\caption{Estimated overall ATT by different methods for justified envy in the NCEE dataset. Standard errors (SE), in parentheses, are clustered at the province level (27 clusters), and $P$-values use the normal approximation; $^{*}P<0.05$, $^{**}P<0.01$, $^{***}P<0.001$.} \label{tab:data1}
\begin{tabular}{lcccccc}
\toprule
Outcome & \multicolumn{3}{c}{BS ($\times1000$)} & \multicolumn{3}{c}{BP ($\times1000$)} \\
\cmidrule(lr){2-4} \cmidrule(lr){5-7}
& ATT & (SE) & $P$ & ATT & (SE) & $P$ \\
\midrule
TWFE & -0.733 & (0.271) & 0.007** & -2.462 & (0.705) & $<$0.001*** \\
DRnt & -0.801 & (0.436) & 0.066 & -2.902 & (0.638) & $<$0.001*** \\
DRny & -0.675 & (0.443) & 0.128 & -2.759 & (0.615) & $<$0.001*** \\
CSX & -0.479 & (0.443) & 0.279 & -2.781 & (0.606) & $<$0.001*** \\
LD & -0.743 & (0.330) & 0.024* & -2.761 & (0.778) & $<$0.001*** \\
AIPW & -0.605 & (0.302) & 0.045* & -2.759 & (0.819) & $<$0.001*** \\
AIVW & -0.513 & (0.298) & 0.085 & -2.696 & (0.818) & $<$0.001*** \\
\midrule
Outcome & \multicolumn{3}{c}{TE ($\times1000$)} & \multicolumn{3}{c}{BT ($\times1000$)} \\
\cmidrule(lr){2-4} \cmidrule(lr){5-7}
& ATT & (SE) & $P$ & ATT & (SE) & $P$ \\
\midrule
TWFE & -3.516 & (0.935) & $<$0.001*** & -4.531 & (1.392) & 0.001** \\
DRnt & -4.068 & (0.838) & $<$0.001*** & -5.027 & (1.450) & $<$0.001*** \\
DRny & -3.826 & (0.809) & $<$0.001*** & -4.668 & (1.376) & $<$0.001*** \\
CSX & -3.909 & (0.912) & $<$0.001*** & -5.385 & (1.229) & $<$0.001*** \\
LD & -3.841 & (1.058) & $<$0.001*** & -4.894 & (1.424) & $<$0.001*** \\
AIPW & -3.822 & (1.040) & $<$0.001*** & -4.680 & (1.388) & $<$0.001*** \\
AIVW & -3.775 & (1.047) & $<$0.001*** & -4.992 & (1.452) & $<$0.001*** \\
\bottomrule
\end{tabular}
\end{table}

\begin{figure}
\centering
\includegraphics[width=0.9\textwidth]{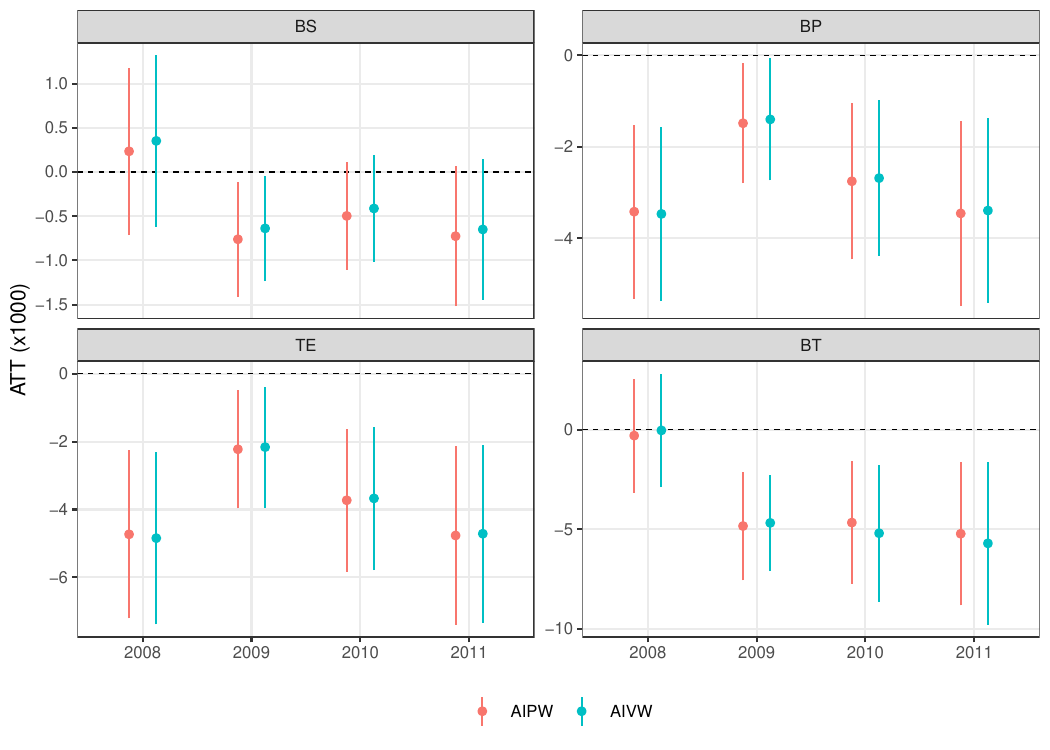}
\caption{Estimated periodwise ATTs for justified envy based on AIPW and AIVW in the NCEE dataset, with 95\% confidence intervals based on standard errors clustered at the province level. The estimates for 2008 rest on the three provinces that adopted in that year.} \label{fig3}
\end{figure}

Figure \ref{fig3} shows the periodwise ATTs of AIPW and AIVW, which are similar. In 2008, they rest on the three provinces that adopted in that year; the estimates for BS and BT are close to zero and not significant, whereas those for BP and TE are negative and significant. From 2009 onward, the estimates are negative for all four outcomes; for BP, TE, and BT, they are significant in every year and larger in magnitude in 2011 than in 2009, whereas the estimate for BS is significant only in 2009. One interpretation is that students and parents learned from the experience of other provinces and improved their applications over time. The groupwise and dynamic ATTs (Supplementary Material B.2) show no clear pattern.

Supplementary Material B reports sensitivity analyses. With standardized justified envy, the observed envy divided by its maximum possible value in each game (B.3), all methods give significant negative overall ATTs. If we adjust only for the baseline covariates (B.4), all give significant negative estimates for BP, TE, and BT, but most are insignificant for BS.

\section{Discussion} \label{sec6}

We proposed doubly robust AIPW and AIVW estimators for staggered DiD under a period-by-period parallel trends assumption with time-varying covariates. Relative to estimators that condition on baseline covariates, they remain consistent when parallel trends hold only given the time-varying covariates and, in our simulations, are more precise when these covariates predict the untreated trend. Relative to a long-difference estimator that conditions on the time-varying covariates at the endpoints, they are preferable when the untreated change in each period depends on the covariates of that period, whereas the long-difference estimator can be more precise when the untreated outcome depends on the covariates only through their current values. The choice between the two should rest on which parallel trends assumption is plausible.

We recommend AIPW as the default. It requires only an outcome regression and a propensity score, and its overall ATT is a weighted sum of outcome-regression residuals (equation \eqref{IF_tau_H}). The gain from AIVW depends on the serial dependence of the errors and on group differences in conditional variance. AIVW can be less precise than AIPW when errors are independent in levels, and its variance model adds finite-sample variability \citep{wang2023model}. We suggest reporting AIPW as the primary estimate and using the $F$-test of Section \ref{sec:linear}, together with the autocorrelation of residual changes, which is near $-0.5$ when errors are independent in levels and near zero when they are independent in increments, to decide whether to report AIVW as a robustness check.

Neither AIVW nor AIPW is efficient in general. The EIF under Assumption \ref{ass3} has not been derived; \citet{chen2025efficient} derive it under parallel trends conditional on baseline covariates \citep[see also][]{chamberlain1992efficiency}. Possible routes are to restrict attention to a class of estimators, such as linear ones, or to strengthen the assumption, for example to parallel trends conditional on the full history \citep{shahn2022structural}, which, as Remark \ref{rmk1} notes, replaces Assumptions \ref{ass2}--\ref{ass3} with an assumption on a treatment-affected history.

The period-by-period assumption cannot be tested against the long-difference assumption. Assumption \ref{ass2} cannot be tested either; in our application it rests on an institutional argument, but two provinces changed their preference-timing rule in the year they adopted the parallel mechanism. Conditioning on $X_k$ in every period requires positivity in every period, which becomes more demanding as the number of not-yet-treated groups shrinks. With few clusters, as in our application, normal-approximation inference may be anticonservative.

\section*{Supplementary Material}
The online supplementary material includes R code, technical details, additional simulation results, and additional results from data analysis.

\section*{Data Availability Statement}
The R code is available in the online supplementary material. The data supporting the findings of this work are available from the corresponding author upon request.

\section*{Conflict of interest}
The authors declare no conflict of interest.

\newpage

\includepdf[pages=1-31]{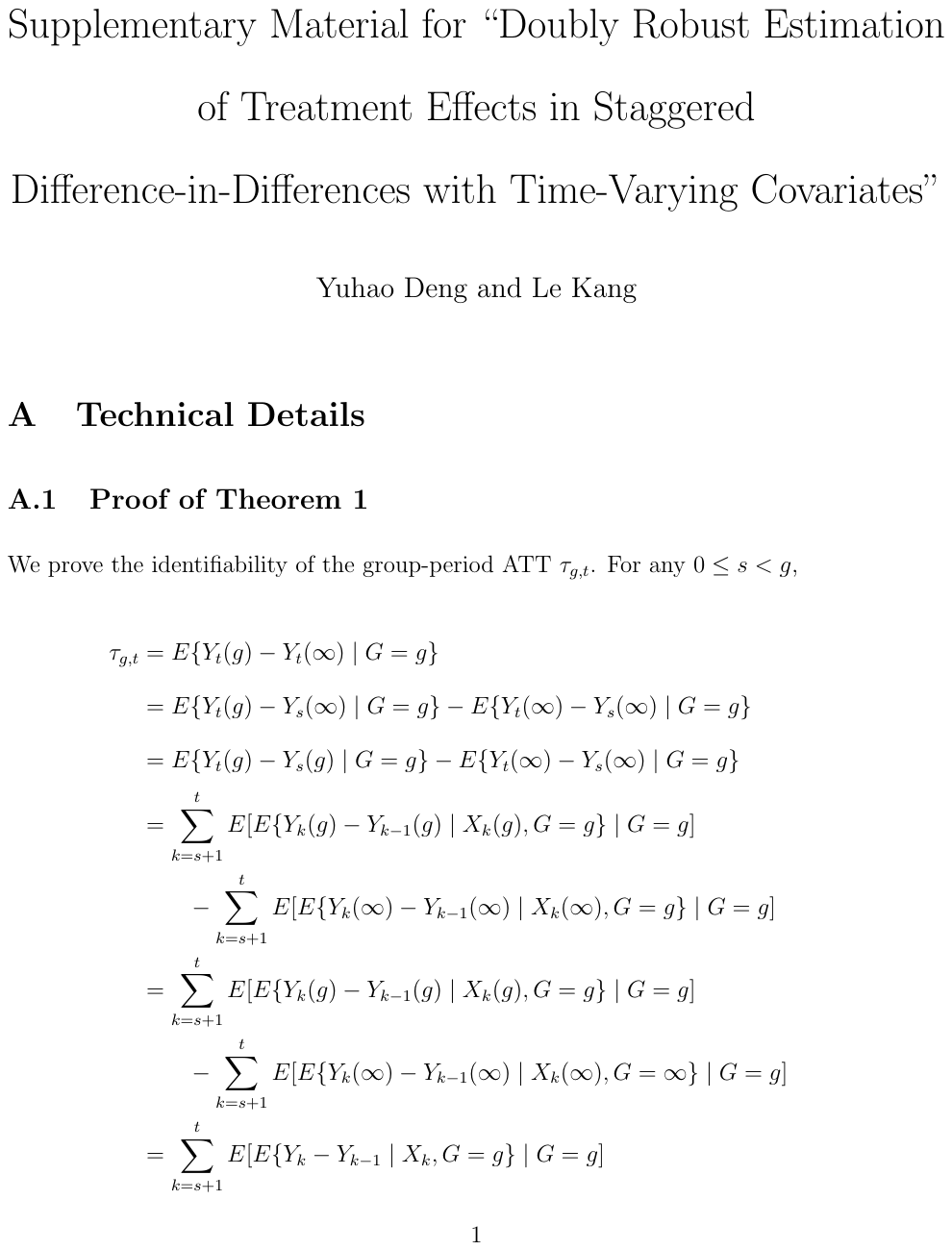}


\begin{thebibliography}{}

\bibitem[Abdulkadiro{\u{g}}lu and S{\"o}nmez, 2003]{abdulkadirouglu2003school}
Abdulkadiro{\u{g}}lu, A. and S{\"o}nmez, T. (2003).
\newblock School choice: A mechanism design approach.
\newblock {\em American Economic Review}, 93(3):729--747.

\bibitem[Athey and Imbens, 2022]{athey2022design}
Athey, S. and Imbens, G.~W. (2022).
\newblock Design-based analysis in difference-in-differences settings with
  staggered adoption.
\newblock {\em Journal of Econometrics}, 226(1):62--79.

\bibitem[Athey et~al., 2019]{athey2019generalized}
Athey, S., Tibshirani, J., and Wager, S. (2019).
\newblock Generalized random forests.
\newblock {\em The Annals of Statistics}, 47(2):1148--1178.

\bibitem[Balinski and S{\"o}nmez, 1999]{balinski1999tale}
Balinski, M. and S{\"o}nmez, T. (1999).
\newblock A tale of two mechanisms: student placement.
\newblock {\em Journal of Economic Theory}, 84(1):73--94.

\bibitem[Caetano and Callaway, 2024]{caetano2024difference}
Caetano, C. and Callaway, B. (2024).
\newblock Difference-in-differences when parallel trends holds conditional on
  covariates.
\newblock {\em arXiv preprint arXiv:2406.15288}.

\bibitem[Caetano et~al., 2024]{caetano2022difference}
Caetano, C., Callaway, B., Payne, S., and Rodrigues, H.~S. (2024).
\newblock Difference-in-differences with time-varying covariates.
\newblock {\em arXiv preprint arXiv:2202.02903v3}.

\bibitem[Callaway and Sant'Anna, 2021]{callaway2021difference}
Callaway, B. and Sant'Anna, P.~H. (2021).
\newblock Difference-in-differences with multiple time periods.
\newblock {\em Journal of Econometrics}, 225(2):200--230.

\bibitem[Chamberlain, 1992]{chamberlain1992efficiency}
Chamberlain, G. (1992).
\newblock Efficiency bounds for semiparametric regression.
\newblock {\em Econometrica: Journal of the Econometric Society}, pages
  567--596.

\bibitem[Chen et~al., 2025]{chen2025efficient}
Chen, X., Sant'Anna, P.~H., and Xie, H. (2025).
\newblock Efficient difference-in-differences and event study estimators.
\newblock {\em arXiv preprint arXiv:2506.17729}.

\bibitem[Chen et~al., 2020]{chen2020empirical}
Chen, Y., Jiang, M., and Kesten, O. (2020).
\newblock An empirical evaluation of chinese college admissions reforms through
  a natural experiment.
\newblock {\em Proceedings of the National Academy of Sciences},
  117(50):31696--31705.

\bibitem[Chernozhukov et~al., 2018]{chernozhukov2018double}
Chernozhukov, V., Chetverikov, D., Demirer, M., Duflo, E., Hansen, C., Newey,
  W., and Robins, J. (2018).
\newblock Double/debiased machine learning for treatment and structural
  parameters.

\bibitem[Chernozhukov et~al., 2022]{chernozhukov2022locally}
Chernozhukov, V., Escanciano, J.~C., Ichimura, H., Newey, W.~K., and Robins,
  J.~M. (2022).
\newblock Locally robust semiparametric estimation.
\newblock {\em Econometrica}, 90(4):1501--1535.

\bibitem[De~Chaisemartin and d’Haultfoeuille, 2020]{de2020two}
De~Chaisemartin, C. and d’Haultfoeuille, X. (2020).
\newblock Two-way fixed effects estimators with heterogeneous treatment
  effects.
\newblock {\em American Economic Review}, 110(9):2964--2996.

\bibitem[Goodman-Bacon, 2021]{goodman2021difference}
Goodman-Bacon, A. (2021).
\newblock Difference-in-differences with variation in treatment timing.
\newblock {\em Journal of Econometrics}, 225(2):254--277.

\bibitem[Ha et~al., 2020]{ha2020college}
Ha, W., Kang, L., and Song, Y. (2020).
\newblock College matching mechanisms and matching stability: Evidence from a
  natural experiment in china.
\newblock {\em Journal of Economic Behavior \& Organization}, 175:206--226.

\bibitem[Kamada and Kojima, 2024]{kamada2024fair}
Kamada, Y. and Kojima, F. (2024).
\newblock Fair matching under constraints: Theory and applications.
\newblock {\em Review of Economic Studies}, 91(2):1162--1199.

\bibitem[Kang et~al., 2020]{kang2020matching}
Kang, L., Ha, W., Song, Y., and Zhou, S. (2020).
\newblock Matching mechanisms, justified envy, and college admissions outcomes.
\newblock {\em Justified Envy, and College Admissions Outcomes (July 22,
  2020)}.

\bibitem[Kuchibhotla and Patra, 2020]{kuchibhotla2020efficient}
Kuchibhotla, A.~K. and Patra, R.~K. (2020).
\newblock Efficient estimation in single index models through smoothing
  splines.
\newblock {\em Bernoulli}, 26(2):1587--1618.

\bibitem[Lechner, 2011]{lechner2011estimation}
Lechner, M. (2011).
\newblock The estimation of causal effects by difference-in-difference methods.
\newblock {\em Foundations and Trends in Econometrics}, 4(3):165--224.

\bibitem[Lee et~al., 2016]{lee2016comparison}
Lee, C.~H., Cook, S., Lee, J.~S., and Han, B. (2016).
\newblock Comparison of two meta-analysis methods: inverse-variance-weighted
  average and weighted sum of z-scores.
\newblock {\em Genomics \& Informatics}, 14(4):173.

\bibitem[Lu et~al., 2007]{lu2007estimation}
Lu, M., Zhang, Y., and Huang, J. (2007).
\newblock Estimation of the mean function with panel count data using monotone
  polynomial splines.
\newblock {\em Biometrika}, 94(3):705--718.

\bibitem[Marcus and Sant’Anna, 2021]{marcus2021role}
Marcus, M. and Sant’Anna, P.~H. (2021).
\newblock The role of parallel trends in event study settings: An application
  to environmental economics.
\newblock {\em Journal of the Association of Environmental and Resource
  Economists}, 8(2):235--275.

\bibitem[Martinez~Taboada et~al., 2023]{martinez2023efficient}
Martinez~Taboada, D., Ramdas, A., and Kennedy, E. (2023).
\newblock An efficient doubly-robust test for the kernel treatment effect.
\newblock {\em Advances in Neural Information Processing Systems},
  36:59924--59952.

\bibitem[Shahn et~al., 2022]{shahn2022structural}
Shahn, Z., Dukes, O., Richardson, D., Tchetgen, E.~T., and Robins, J. (2022).
\newblock Structural nested mean models under parallel trends assumptions.
\newblock {\em arXiv preprint arXiv:2204.10291}.

\bibitem[Sun and Abraham, 2021]{sun2021estimating}
Sun, L. and Abraham, S. (2021).
\newblock Estimating dynamic treatment effects in event studies with
  heterogeneous treatment effects.
\newblock {\em Journal of Econometrics}, 225(2):175--199.

\bibitem[Tsiatis, 2006]{tsiatis2006semiparametric}
Tsiatis, A.~A. (2006).
\newblock {\em Semiparametric theory and missing data}, volume~4.
\newblock Springer.

\bibitem[{Van der Vaart}, 2000]{van2000asymptotic}
{Van der Vaart}, A.~W. (2000).
\newblock {\em Asymptotic statistics}, volume~3.
\newblock Cambridge University Press.

\bibitem[{van der vaart} and Wellner, 2013]{van2013weak}
{van der vaart}, A.~W. and Wellner, J. (2013).
\newblock {\em Weak Convergence and Empirical Processes: With Applications to
  Statistics}.
\newblock Springer Series in Statistics. Springer New York.

\bibitem[Wang et~al., 2023]{wang2023model}
Wang, B., Susukida, R., Mojtabai, R., Amin-Esmaeili, M., and Rosenblum, M.
  (2023).
\newblock Model-robust inference for clinical trials that improve precision by
  stratified randomization and covariate adjustment.
\newblock {\em Journal of the American Statistical Association},
  118(542):1152--1163.

\bibitem[Wooldridge, 2025]{wooldridge2025two}
Wooldridge, J.~M. (2025).
\newblock Two-way fixed effects, the two-way mundlak regression, and
  difference-in-differences estimators.
\newblock {\em Empirical Economics}, 69(5):2545--2587.

\end{thebibliography}
\end{document}